\documentclass[useAMS,usenatbib,usegraphicx]{mn2e}

\title[]{Non-Gaussianity in the Very Small Array CMB maps with
Smooth-Goodness-of-fit tests}

\author[J.A. Rubi\~no-Mart\'in et al.]{
Jos\'e Alberto Rubi\~no-Mart\'in$^1\thanks{Email: jalberto@iac.es}$,
 Antonio M. Aliaga$^{2}$, 
 R. B. Barreiro$^{2}$,
\newauthor
 Richard A. Battye$^3$,
 Pedro Carreira$^3$, 
 Kieran Cleary$^3 \thanks{Present address: Jet Propulsion Laboratory, 
 4800 Oak Grove Dr., MS 264-767, Pasadena, CA 91109}$, 
 Rod D. Davies$^3$, 
\newauthor  
 Richard J. Davis$^3$,
 Clive Dickinson$^3 \thanks{Present address: California Institute of Technology, Dept. of
  Astronomy, MS 105-24, 1200 E. California Blvd., Pasadena, CA 91125.}$, 
 Ricardo G\'enova-Santos$^1$,
 Keith Grainge$^4$, 
\newauthor
 Carlos M. Guti{\'e}rrez$^1$,  
 Yaser A. Hafez$^3$,
 Michael P. Hobson$^4$,  
 Michael E. Jones$^4\thanks{Present address: Astrophysics Group, Denys Wilkinson Building,
  University of Oxford.}$, 
\newauthor
 R\"udiger Kneissl$^4$, 
 Katy Lancaster$^4$, 
 Anthony Lasenby$^4$,  
 J. P. Leahy$^3$,
 Klaus Maisinger$^4$,
\newauthor
 Enrique Mart\'inez-Gonz\'alez$^2$, 
 Guy G. Pooley$^4$, 
 Nutan Rajguru$^4$, 
 Rafael Rebolo$^{1,5}$, 
\newauthor
 Jos\'e Luis Sanz$^2$, 
 Richard D.E. Saunders$^4$, 
 Richard S. Savage$^4\thanks{Present address: Astronomy Centre, University of
   Sussex, UK.}$,
 Anna Scaife$^4$,
\newauthor  
 Paul Scott$^4$,
 An\v ze Slosar$^4\thanks{Present address: Faculty of Mathematics \& Physics, 
University of Ljubljana, 1000 Ljubljana, Slovenia.}$,
 Angela C. Taylor$^4\thanks{Present address: Astrophysics Group, Denys Wilkinson Building, University of Oxford.}$,  
 David Titterington$^4$,
\newauthor  
 Elizabeth Waldram$^4$, 
 Robert A. Watson$^3\thanks{Present address: Instituto de Astrof{\'{\i}}sica de
Canarias, 38200 La Laguna, Tenerife, Spain.}$
\\
  $^1$Instituto de Astrof{\'{\i}}sica de Canarias, 38200 La Laguna,
  Tenerife, Spain.\\ 
  $^2$IFCA, CSIC-Univ. de Cantabria, Avda. los Castros, s/n, E-39005 Santander, Spain \\
  $^3$University of Manchester, Jodrell Bank Observatory, UK.\\
  $^4$Astrophysics Group, Cavendish Laboratory, University of Cambridge, UK.\\
  $^{5}$Consejo Superior de Investigaciones Cient\'\i ficas, Spain.}

\begin{document}

\date{Accepted Received In original form}

\pagerange{\pageref{firstpage}--\pageref{lastpage}} \pubyear{2005}

\maketitle

\label{firstpage}

\begin{abstract}
We have used the Rayner \& Best (1989) smooth tests 
of goodness-of-fit to study the
Gaussianity of the Very Small Array (VSA) data.
These tests are designed to be sensitive to the presence 
of `smooth' deviations from a given
distribution, and are applied to the data transformed into 
normalised signal-to-noise eigenmodes.
In a previous work, they have been already adapted and 
applied to simulated observations of interferometric experiments. 
In this paper, we extend the practical 
implementation of the method to deal with
mosaiced observations, by introducing the Arnoldi algorithm. 
This method permits us 
to solve large eigenvalue problems with low computational cost.

Out of the $41$ published VSA individual pointings dedicated to 
cosmological (CMB) observations, $37$ are
found to be consistent with Gaussianity, 
whereas four pointings show deviations from Gaussianity. 
In two of them, these deviations can be explained as residual
systematic effects of a few visibility points which, when corrected, 
have a negligible impact on the angular power spectrum.   
The non-Gaussianity found in the other two (adjacent) pointings
seems to be associated to a local deviation 
of the power spectrum of these fields with respect
to the common power spectrum of the complete data set,  
at angular scales of the third acoustic peak ($\ell = 700-900$). 
No evidence of residual systematics is found in this case, 
and unsubstracted point sources are not a plausible explanation either. 
If those visibilities are 
removed, the differences of the new power spectrum with respect to the 
published one only affect three bins. 
A cosmological analysis based on this new VSA power spectrum alone shows 
no differences in the parameter constraints with respect to our published
results, except for the physical baryon density, which decreases 
by $10$~percent. 

Finally, the method has been also used to analyse the VSA observations
in the Corona Borealis supercluster region. 
Our method finds a clear deviation ($99.82\%$) with respect to
Gaussianity in the second-order moment of the distribution, 
and which can not be explained as systematic
effects. A detailed study shows that the non-Gaussianity is produced
in scales of $\ell \approx 500$, and that 
this deviation is intrinsic to the data (in
the sense that can not be explained in terms of a Gaussian field
with a different power spectrum). 
This result is consistent with the Gaussianity studies in the Corona
Borealis data
presented in G\'enova-Santos et al. (2005), 
which show a strong decrement which cannot be explained as primordial CMB.

\end{abstract}


\section{Introduction}

The study of the Gaussianity of the primordial density
fluctuations is a very important tool in constraining theories of
structure formation. Inside the inflationary paradigm, there is
a huge number of theories (see \citet{bartolo} for a recent 
review on the subject), each one predicting different non-Gaussian
signatures. Thus, any detection of non-Gaussianity 
would help to discriminate among these scenarios for the generation
of cosmological perturbations.
Because of this reason, the study of the Gaussianity of Cosmic
Microwave Background (CMB) maps is becoming of major importance 
in modern cosmology. In particular, since the publication of the
first year WMAP results \citep{bennett}, several groups have
tested the non-Gaussian nature of those maps using a wide
set of techniques 
\citep{komatsu,chiang,eriksen04a,eriksen04b,vielva,park,cruz}.

Furthermore, there are other reasons showing the importance of the 
study of the Gaussianity of the CMB.
The majority of the inflationary models predict the primordial 
non-Gaussian signal to be smaller than the contribution from secondary
effects such as gravitational lensing, reionization, Sunyaev-Zel'dovich
effect, or the contribution of local foregrounds or unresolved
point sources in the maps. 
Thus, tools to test Gaussianity could be used
to trace the presence of these foregrounds.
For example, the analysis of the 
WMAP data using the bispectrum allowed \cite{komatsu} to perform
estimates of the source number counts of unresolved sources in 
the 41~GHz channel (see also \cite{2005ApJ...621....1G}).

In addition, systematic effects may produce spurious detections of 
non-Gaussianities, so non-Gaussian methods could help in 
characterizing the properties of a given experiment 
(e.g. \cite{2000ApJ...533..575B}).

The Gaussianity of the VSA data was already examined using several 
methods in two separate papers \citep{savage,smith}, 
which were based on the data
presented in \cite{vsa2} and \cite{vsa5}.
In \cite{savage}, a selection of non-Gaussianity
tests are applied to the data. Most of these tests are based on real-space 
statistics and are applied to the maximum-entropy reconstruction of the
regions observed by the instrument.
In \cite{smith}, the analysis is devoted to the study of the 
bispectrum of the VSA data, showing how this statistic can be obtained
in the case of interferometric experiments.

In this paper, we present the results of a Gaussianity analysis 
of the complete set of observations of  
the Very Small Array (VSA) dedicated to measure the CMB power spectrum
(see \cite{vsa7} and references therein), as well as an analysis of the data from the
Corona-Borealis supercluster survey presented in \cite{ricardo}. 
Here, we will complement the previous Gaussianity studies of the VSA
data by considering a different family of methods, called 
the Smooth Tests of Goodness of Fit (STGOF).

In section 2, we give a brief overview of the VSA experiment. 
In Section 3, we review the Smooth Tests of Goodness of Fit methods, and
how these method can be adapted to the study of the Gaussianity
of interferometric experiments.
Section 4 describes how these methods can be further adapted to
deal with large datasets or mosaiced observations.
Section 5 presents the calibration of the method using Gaussian
simulations of mosaiced observations with the VSA. 
Section 6 presents the results of our analysis, and finally conclusions
are presented in section 7.

\section{The Very Small Array}

The VSA is a 14-element heterodyne interferometer sited at the 
Teide Observatory (Tenerife). The instrument is designed to image the
CMB on scales going from $2\degr$ to 10\arcmin, and operates
at frequencies between $26$ and $36$~GHz with a $1.5$~GHz bandwidth
and a system temperature of $\sim 30$~K. 
The VSA has observed in two configurations of antennas.
The first one is the so-called 
'compact configuration', which covers the multipole range 
$\ell \sim 150-900$ with 
a primary beam of $4.6\degr$-FWHM at 34.1~GHz. 
This configuration was used during the first observing season (September 2000-
September 2001). The results of this campaign are presented in
\cite{vsa1,vsa2,vsa3} and \cite{vsa4}.  

The second one, the 'extended configuration', provided 
observations up to $\ell = 1500$  
with a primary beam of $2.10\degr$-FWHM (at $33$~GHz) 
and an angular resolution
of 11~arcmin during two separate campaigns. 
Those results were presented in two separate sets of papers: 
\cite{vsa5,vsa6} for the second season of observations (September 2001 -
April 2002); and \cite{vsa7,vsa8} for the third one (April 2002-
January 2003), where we obtained maps both at 34.1~GHz and at 33~GHz.
With this extended configuration, we have 
  obtained maps of a complete, 
  X-ray flux-limited sample of seven clusters with 
  redshifts $z<0.1$ \citep{lancaster}. 
We have also produced imaging at $33$~GHz of the 
Corona-Borealis supercluster \citep{ricardo},  
with the aim of searching for
Sunyaev-Zel'dovich detections 
from a possible extended signal due to diffuse warm/hot gas. 
As shown in that paper, we found a strong decrement near the centre
of the supercluster, which can not be associated either with primordial
CMB fluctuations or with a SZ effect from a known cluster of galaxies in 
the region. Therefore, we shall consider these data in our
Gaussianity analysis as well.

\begin{table*}
\caption{Summary of the VSA observations dedicated to cosmological
studies, and which have been analysed in this paper.
We present the names for individual pointings contributing to one of the $7$ 
VSA mosaics, separating the observations according to the three VSA 
campaigns. The central coordinates, 
integration times and maps for each of the
individual pointings can be found in the three specified references. }
\label{tab:data}
\centering
\begin{tabular}{@{}cccc}
\hline
\hline
Mosaic  & Compact       & Extended       & Extended II \\
        & \citep{vsa2}   & \citep{vsa5}    & \citep{vsa7}   \\
\hline
VSA1 & 1, 1A, 1B & 1E, 1F, 1G & 1H, 1J, 1K, 1L \\
VSA2 & 2, 2-OFF  & 2E, 2F, 2G & 2H, 2J, 2K, 2L\\
VSA3 & 3, 3A, 3B & 3E, 3F, 3G & 3H, 3J, 3K, 3L\\
VSA5 &  -        &     -      & 5E, 5F, 5G \\
VSA6 &  -        &     -      & 6E, 6F, 6G \\
VSA7 &  -        &     -      & 7E, 7F, 7G \\
VSA8 &  -        &     -      & 8E, 8F, 8G \\
\hline
\hline
\end{tabular}
\end{table*}

In Table~\ref{tab:data} we summarise the whole set of observations 
obtained with the VSA and used for cosmological studies, both with 
its compact and extended configurations. 
The full dataset comprises $8$ fields observed with the compact array, and 
$33$ fields observed with the extended one. This dataset can be arranged
into seven separate (not overlapping) regions on sky, each of them 
obtained using mosaicing of individual pointings. 
Each mosaiced field is labeled as ``VSA'' plus a number. Within each mosaic, 
the names of the individual pointings are denoted by either no suffix, or
the suffixes A, B, -OFF, E, F, G, H, J, K and L. 
Detailed information about the fields (central coordinates, integration times 
and maps) can be found in the indicated references. 
All these regions were carefully chosen to minimise contamination 
from Galactic emission and bright radio sources. 
Further details of the residual contamination in the maps
can be found in \cite{vsa7}, and details about the VSA observational
technique can be found in \cite{vsa1}.

Regarding the Corona Borealis observations, the core of the supercluster
is imaged with a $9$ pointings mosaic, 
and we have two additional pointings outside
this region to map two supercluster members which lie far from the
optical centre of the supercluster.   
The total area covered is $\sim 24$ deg$^2$ 
with an angular resolution of 11 arcmin and a sensitivity of 12 mJy/beam. 
Detailed information about the fields (central coordinates, integration times 
and maps) can be found in \cite{ricardo}.

\subsection{Interferometer measurements}

For observations of small patches of sky, we can adopt the flat-sky 
approximation and use Fourier analysis instead of the spherical harmonic
expansion for the temperature field.  
In this limit, the {\sl complex visibility} (which gives the response of an
interferometer observing at frequency $\nu$) can be written as

\begin{equation}
V( \bmath{u} ,\nu)=\int P(\hat{\bmath{x}} , \nu) B(\hat{\bmath{x}}, \nu)
\exp{(i 2 \pi  \bmath{u} \cdot \hat{\bmath{x}} )} 
d\hat{\bmath{x}}
\label{eq:visibility}
\end{equation}
where $\hat{\bmath{x}}$ is the angular position of the 
observed point on the sky; $\bmath{u}$ is the 
baseline vector in units of the wavelength of the observed radiation (so  
$2\pi \bmath{u}$ is the Fourier mode); 
$P(\hat{\bmath{x}} ,\nu)$ is the primary beam of the antennas 
(normalised to unity at its peak); 
and $B(\hat{\bmath{x}} ,\nu)$ is the brightness distribution on the sky. 
For the case of CMB observations, this brightness can be expressed in
terms of the equivalent thermodynamic temperature fluctuations 
($\Delta T (\hat{\bmath{x}} )$) as

\begin{equation}
B(\hat{\bmath{x}} , \nu) \approx 
\frac{\partial B_\nu(T)}{\partial T} \Bigg |_{T=T_0} 
\Delta T (\hat{\bmath{x}} )
\end{equation}
where $B_\nu(T)$ is the Planck function, and the mean
temperature of the CMB is given by $T_0=2.726$~K \citep{1994ApJ...420..439M}.

By inserting the Fourier decomposition of the sky brightness in 
equation~\ref{eq:visibility}, we find that 
an interferometer measures the convolution of sky Fourier modes
with the aperture function (Fourier transform of the primary beam), 
sampling at those points given by
the projection of the baselines on the sky plane.  

We note that the previous equation does not take into account
the contribution of instrumental noise. Thus, for a realistic
instrument observing at a frequency $\nu$, the {\it i}th 
baseline $\bmath{u}_i$ of the interferometer 
will measure the following quantity 
\[
d(\bmath{u}_i, \nu) = V(\bmath{u}_i, \nu)  + n(\bmath{u}_i, \nu) 
\]
where $n( \bmath{u}_i, \nu)$ stands for the instrumental noise
on the $\bmath{u}_i$ visibility.

Let $N$ be the total number of complex visibilities observed by an
interferometer. 
Then, the complete set of observed visibilities
will be noted as the following vector with $N_d = 2N$ elements
\[
\bmath{d} = \{ \Re [ d(\bmath{u}_1, \nu_1) ], \ldots ,
\Re [ d(\bmath{u}_N, \nu_N) ], 
\Im [ d(\bmath{u}_1, \nu_1) ], 
\]
\[
\qquad \ldots , \Im [ d(\bmath{u}_N, \nu_N) ]  \}
\]
where the label $\Re$ ($\Im$) stands for the real (imaginary) part of the
complex number. 
We must note that in this equation,  
we explicitly differentiate the observing frequency 
for each observed sample because, in general, we could combine data taken 
at different frequencies with the same instrument (this is indeed the
case of the VSA, in which we have observations at two different
frequencies, 33~GHz and 34.1~GHz).

From here, we can also define the vector for the
sky signal, and the vector for the noise in the same way, so  
we have $\bmath{d} = \bmath{V} + \bmath{n}$. 
Assuming that there are no correlations between the sky signal and the
noise, the covariance matrix for this set of observations can 
be written as
\[
\bmath{C} = < \bmath{d} \bmath{d}^t > = \bmath{S} + \bmath{N}
\] 
where $\bmath{S}$ and $\bmath{N}$ are the covariance matrices of the
contributions from the sky signal and the noise, respectively. 
In our analysis, we shall take this noise covariance matrix as diagonal,
as is the case of the VSA data (e.g. \cite{vsa7}).

The covariance matrix for the CMB component ($\bmath{S}$)
can be computed analytically using the equations 
presented in \cite{hobson02},  
both for the case of a single or mosaiced observations.
If the primary beam of the interferometer horns is symmetric respect to 
inversion through the origin (as it is the case of the VSA experiment, where
the primary beam can be modelled to a good approximation by an spherical 
Gaussian function), 
then the aperture function is real. As a consequence, 
for the case of single-field observations the covariance matrix is
block-diagonal (i.e. the real and imaginary parts of the 
visibilities are uncorrelated). 
Note that for mosaiced observations, this is not true in general.

%
%
\section{Goodness-of-fit statistics applied to interferometers}

In this section we summarise some aspects of the smooth goodness-of-fit
tests applied to CMB interferometers. For a more detailed description, 
see \cite[hereafter, A05]{paper1}.

As shown in the previous section, the visibilities observed by
an interferometer are correlated quantities. Therefore, the STGOF 
have to be adapted to deal with these data, because in their original form,
these tests require independent data points. 
As described in A05, this is done following a two-step procedure. 
First, the data $\bmath{d}$ are transformed into
signal-to-noise eigenmodes $\bmath{\xi}$ \citep{bond}, 
and they are normalised. 
After this, the smooth goodness-of-fit tests developed by \cite{rayner} 
can be applied to the normalised eigenmodes, which for the Gaussian case would
be independent.

\subsection{Signal-to-noise eigenmodes}

In a first step, the data are transformed into 
signal-to-noise eigenmodes as explained in A05. 
Every eigenmode has an associated signal-to-noise eigenvalue, in such 
a way that the higher is the value of the eigenvalue the more 
signal-to-noise ratio is associated to the eigenmode.
Thus, this decomposition permits us not only to decorrelate the visibilities, 
but also to select those data points in which the signal contribution 
is dominating over that of the noise.

Let $\bmath{L}_n$ be the {\emph square
root matrix} of the noise correlation matrix (i.e. 
$\bmath{N} =  \bmath{L}_n \bmath{L}_n^t$), and $\bmath{R}$ the
rotation matrix which diagonalizes the matrix 
\begin{equation}
\bmath{A} \equiv \bmath{L}_n^{-1} \bmath{S} \bmath{L}_n^{-t}
\label{eq:A}
\end{equation}
Thus, $\bmath{R}^t \bmath{A} \bmath{R} = \bmath{E}$, where
$\bmath{E} = \mbox{diag}(E_1,\ldots, E_{N_d})$ 
is a diagonal matrix whose diagonal elements are 
the signal-to-noise eigenvalues ($E_i$).
With these definitions, the signal-to-noise eigenmodes are obtained as

\begin{equation}
\bmath{\xi} = \bmath{R}^t \bmath{L}_n^{-1} \bmath{d}
\label{eq:eigenmodes}
\end{equation}
From here, it is easy to show that the covariance matrix associated
to these variables is given by 
$< \bmath{\xi} \bmath{\xi}^t > = \bmath{E} + \bmath{I}_{N_d}$, 
where $\bmath{I}_{N_d}$ is the identity matrix with dimension 
$N_d \times N_d$. 

The normalised signal-to-noise eigenmodes can be defined from here
as $y_i= \xi_i / (E_i+1)^{1/2}$ ($i=1,\ldots, N_d$), and one can
immediately show that these quantities are uncorrelated and
they verify $\langle y_i y_j \rangle = \delta_{ij}$.

For our case of interest (CMB analyses), the important point is
that equation~\ref{eq:eigenmodes} is a linear transformation, so it 
preserves the Gaussianity of the
variables (i.e. if the data $\bmath{d}$ are distributed
following a multi-normal function, then the normalised eigenmodes will
follow a one-dimensional Gaussian distribution $N(0,1)$). 
Thus, one can now apply the STGOF to these transformed variables.

\subsection{Smooth tests of goodness-of-fit}

Let us assume that we have $n$ independent realizations 
$\{x_i\}_{i=1}^{i=n}$ of a statistical variable $x$, 
and we want to test if $x$ has a distribution 
function compatible with $f(x)$ (null hypothesis).
\cite{rayner} proposed some statistics to discriminate between
$f$ and another distribution (alternative hypothesis) which deviates smoothly
from $f$. 
In the case in which $f$ is a Gaussian ($N(0,1)$), it
can be shown that the first four score statistics associated with
the alternative are given by
\begin{equation}
S_k = \sum_{i=1}^{k} U_i^2
\end{equation}
with
\begin{eqnarray}
U_1^2  &=& n ( \hat{\mu}_1)^2          \nonumber\\
U_2^2  &=& n ( \hat{\mu}_2 - 1 )^2 /2                             \nonumber\\
U_3^2  &=& n ( \hat{\mu}_3 - 3 \hat{\mu}_1 )^2 /6                 \nonumber\\
U_4^2  &=& n [(\hat{\mu}_4 - 3 ) - 6 (\hat{\mu}_2 -1) ]^2  /24 
\end{eqnarray}
where $\hat{\mu}_{\alpha}=(\sum_{j=1}^n x_j^{\alpha})/n$ is the estimated
moment of order $\alpha$. It should be noted that this test
is \emph{directional}, i.e. it indicates how the actual
distribution deviates from Gaussianity. 

If the data are drawn from the distribution given by $f$, 
and $n$ is large enough ($n \ga 100$), then it is possible 
to show that the $U_i^2$ quantities
are distributed following a $\chi^2$ with one degree of freedom.
If the data do not follow an $f$ distribution, we expect departures from this
distribution, and this is the way we detect non-Gaussian signals. 

We shall apply these statistics to the normalised eigenmodes described in 
the last subsection. The important point for us is that, as shown in A05, 
we can select subsets of signal-to-noise 
eigenmodes, according to their associated eigenvalue $E_i$.
This will permit us to test if $y_i \sim N(0,1)$ (that is,
our null hypothesis is that $f$ is a Gaussian function).

\section{Extending the method to mosaiced fields. The Arnoldi algorithm}

The method we have applied in the previous sections only uses a few eigenmodes 
(the ones whose eigenvalue is large enough), but the correlation matrices are 
relatively large (matrices from 
5000$\times$5000 to 10000$\times$10000 for the mosaic analysis). This implies 
a big computational cost in the diagonalization of the matrices. 
Thus, we pose the following question: is it possible to reduce the dimension 
of the correlation matrix to calculate only those eigenmodes and 
eigenvalues we need? 

There are several numerical methods for solving eigenvalue problems
with large matrices. We shall consider here one particular class of 
methods, named the Krylov subspace methods, and in particular 
the Arnoldi algorithm. 
This method can be applied to general non-Hermitian matrices, although
we will focus here in the application to our problem, in which the covariance
matrix is real and symmetric, and has a sparse structure. 
Our development is based on \cite{saad} (see also \citet{aliaga_tesis}).

\subsection{The Arnoldi method}

This procedure was introduced as a means of reducing a dense
matrix into Hessenberg form (lower triangular matrix). However, 
the important point for us is that this method was shown to 
be a good technique for approximating the eigenvalues of large 
sparse matrices. 

The basis of the method is as follows. 
We start from our sparse matrix $\bmath{A}$ given in equation~\ref{eq:A}, which
has dimensions $N_d \times N_d$. 
Then, we want to built a new matrix $\bmath{H}$ 
with dimensions $m\times m$, in such a way that $m < N_d$ 
and the eigenvalues/eigenvectors of $\bmath{H}$ should be approximations to 
the eigenvalues/eigenvectors of $\bmath{A}$ 
(it is obvious that we will not recover all the eigenvalues, 
but only $m$ at the most).

The algorithm produces a set of vectors,  
$\{ \bmath{q}_1, \ldots, \bmath{q}_m \}$, which form an
orthonormal basis of the subspace linear span of 
$\{ \bmath{q}_1, \bmath{A} \bmath{q}_1, \ldots, \bmath{A}^{m-1} \bmath{q}_1\}$.
One variant of this algorithm for real and symmetric matrices will
be presented below.  
From these vectors, we can build the following ($N_d \times m$)-matrix
\[
\bmath{Q}_{ij} \equiv (\bmath{q}_j)_{i}, \qquad i=1,...,N_d \; ; \; j=1,...,m 
\]
with $(\bmath{q}_j)_{i}$ the $i$th component of the $j$th vector.
The matrix in which we are interested can be derived as
\begin{equation}
\bmath{H} = \bmath{Q}^\dag \bmath{A} \bmath{Q}
\label{eq:H}
\end{equation}
where $\dag$ stands for the transpose conjugate matrix\footnote{In our
case, $\bmath{A}$ is real, so we could use transpose instead
of transpose conjugate.}.
The $\bmath{H}$ matrix has an upper triangular form, and can be diagonalized
to find its eigenvalues $E_i^{(H)}$ and associated 
eigenvectors $\bmath{y}_i^{(H)}$. 
From here, we can build the \emph{Ritz approximate eigenvectors} associated
to $E_i^{(H)}$ as $\bmath{e}_i^{(H)} = \bmath{Q} \bmath{y}_i^{(H)}$. 
The important point for us is that a fraction of 
these Ritz eigenvectors are a 
good approximation of the corresponding eigenvectors of $\bmath{A}$, and
at the same time, $E_i^{(H)}$ give a good approximation to
the associated eigenvalue $E_i$. Moreover, the quality of the 
approximation improves as $m$ increases. 

We note that there are simple analytical expressions for 
the residual norm associated to the Ritz eigenvectors. These
expressions can be easily implemented in the algorithm, so 
one controls the quality of the approximations \citep{saad,aliaga_tesis}.

\subsection{Signal-to-noise eigenmodes}

We show now how to use signal-to-noise eigenmodes 
together with the Arnoldi method. 

Let $\bmath{R}^{(H)}$ be the rotation matrix which diagonalizes $\bmath{H}$.
$\bmath{R}^{(H)}$ is constructed in such a way that its $i$th column 
corresponds to the $\bmath{y}_i^{(H)}$ eigenmode defined in 
the previous subsection (i.e. 
$\bmath{R}^{(H)}_{ij} = (\bmath{y}_j^{(H)} )_i$). Then, 
\begin{equation}
(\bmath{R}^{(H)})^t \bmath{H} \bmath{R}^{(H)} = 
\mbox{diag}( E_1^{(H)}, \ldots, E_m^{(H)})
\equiv \bmath{E}^{(H)} 
\label{eq:E_H}
\end{equation}
We now define the matrix $\bmath{T} = \bmath{Q} \bmath{R}^{(H)}$, which
has dimensions $N_d \times m$. Hence, using equations \ref{eq:H} and 
\ref{eq:E_H} we have
\[
\bmath{T}^t \bmath{A} \bmath{T} = \bmath{E}^{(H)} 
\]

Using these matrices, the signal-to-noise eigenmodes can be defined as
\begin{equation}
\bmath{\xi}^{(H)} = \bmath{T}^t \bmath{L}_n^{-1} \bmath{d}
\label{eq:xi_H}
\end{equation}
and the corresponding correlation matrix can be written as
$\langle \bmath{\xi}^{(H)} (\bmath{\xi}^{(H)})^t \rangle 
= \bmath{E}^{(H)} + \bmath{I}_m$. 
The transformed signal and noise vectors are given here 
by $\tilde{\bmath{V}} = \bmath{T}^t \bmath{L}_n^{-1} \bmath{V}$ and
$\tilde{\bmath{n}} = \bmath{T}^t \bmath{L}_n^{-1} \bmath{n}$, such that
$\langle \tilde{\bmath{V}} \tilde{\bmath{V}}^t \rangle = \bmath{E}^{(H)}$ and
$\langle \tilde{\bmath{n}} \tilde{\bmath{n}}^t \rangle = \bmath{I}_m$. 
Therefore, the meaning of the signal-to-noise eigenmodes is preserved, 
and eigenmodes associated to large eigenvalues have more contribution from
the signal than from the noise.

As indicated in the previous subsection, some of the eigenvalues $E_i^{(H)}$
will also be eigenvalues of $\bmath{A}$ 
to a good approximation. Moreover, as we 
shall see below, those eigenvalues which are better approximated will 
correspond to high values of the signal-to-noise eigenvalue. 
Thus, this method will permit to estimate the eigenvalues (and the associated
signal-to-noise eigenmodes) which are of interest for our analysis, but
the dimensionality of the problem will be greatly reduced.
Finally, let us note that with the previous 
definition of signal-to-noise eigenmodes associated to $\bmath{H}$, 
they will directly give a good approximation to the corresponding
eigenmodes for $\bmath{A}$.

\subsection{Relationship between the matrices which diagonalize 
$\bmath{H}$ and $\bmath{A}$}
 
From the definition of the signal-to-noise eigenvectors given
in equation~\ref{eq:xi_H}, it 
is clear that if a pair $E_i^{(H)}$,$\bmath{e}_i^{(H)}$
(an eigenvalue and its associated 
Ritz eigenvector) is a good approximation 
to the corresponding pair for the 
$\bmath{A}$ matrix, then $\xi^{(H)}_i$ will give a 
good approximation to $\xi_i$ as well. However, there is a sign ambiguity
when implementing this algorithm in practice. 
When obtaining the eigenvectors for a given matrix, we impose that
it should be unitary, but an ambiguity in the sign is still present 
(if $\bmath{y}$ is an eigenvector, 
then $-\bmath{y}$ is an eigenvector as well). 

This ambiguity can be avoided by imposing an additional condition, 
for example, that the first component different 
from zero of each eigenvector should be positive (in fact, due to precision
problems, it is better to impose that the component which is the maximum
in absolute value for each eigenvector should be positive). 
With this criterion, then it is clear that the 
statistics of the smooth goodness-of-fit tests of \cite{rayner} 
can be computed from the signal-to-noise eigenmodes 
of $\bmath{H}$, yielding exactly the same values as those computed from
$\bmath{A}$ \emph{if} they only involve those eigenvectors
whose associated eigenvalues are correctly approximated. 
In the next two subsections we present an example of implementation of this
method, and we apply it to the case of simulated VSA observations.

\subsection{Lanczos algorithm}

This is a particular simplification of the Arnoldi algorithm 
for the case when the considered matrix is Hermitian. 
In this case, it can be shown that the $\bmath{H}$ matrix 
is real, tridiagonal and symmetric. 
Thus, the algorithm becomes computationally faster, and fewer variables 
need to be stored in memory. 
The implementation of the algorithm that we have used here is the
following:
 
\begin{enumerate}
\item {\bf Start}. We choose an initial unitary vector $\bmath{q}_1$, and
we define  $H_{01}=0$, $\bmath{q}_0=0$.
\item  {\bf Iterate}. For $j=1,\ldots,m$:
\begin{eqnarray}
\bmath{w}       &=& \bmath{A} \bmath{q}_j - H_{j-1,j} \bmath{q}_{j-1}  \\
H_{jj}          &=& \bmath{q}_j^\dag \bmath{w}        \\
\bmath{w}       &=& \bmath{w} - H_{jj} \bmath{q}_j               \\
H_{j,j+1}       &=& ||\bmath{w}||                      \\
\bmath{q}_{j+1} &=& \bmath{w} / H_{j,j+1}               
\end{eqnarray}
\end{enumerate}
In the previous algorithm, the symbol $||.||$ represents the Euclidean norm of 
a vector, and it is defined as $||\bmath{x}|| \equiv 
(\bmath{x}^\dag \bmath{x})^{1/2}$, where the symbol $\dag$ stands for 
the transpose conjugate operator. 

The above algorithm guarantees that the $\bmath{q}_i$ vectors are orthogonal.
However, when  $m$ is relatively large, the orthonormality of 
the $\bmath{q}_i$ vectors is lost, so they have to be orthonormalised 
as they are calculated (for example using the method of Gram-Schmidt).
We note that if $\bmath{A}$ is sparse (as is our case), then
there are algorithms to optimise the product operation $\bmath{A} \bmath{q}_j$, and
therefore the iterative process can be accelerated (see \citet{saad}).

\section{Calibration of the method with simulated VSA mosaiced observations}

The STGOF have been already applied to simulated observations of 
the VSA in A05, showing the ability of the method to detect non-Gaussian
signals introduced via the Edgeworth expansion, cosmic strings or
$\chi^2$-simulations, when we use realistic noise levels from the 
experiment.
In that paper, the analysis was performed by simulating a single
VSA pointing, and setting the noise levels according to the typical integration
time of the fields.
In order to complete this picture, we present here the calibration of
the method by using simulated Gaussian mosaiced observations.
There is no difference in practice between analysing single and mosaiced
fields, so the results will be similar to those found in A05. 
However, this study will permit us to test the software
for computing covariance matrices in mosaiced observations. 
This software evaluates the covariance matrix from a given power spectrum, 
using the equations presented in \cite{hobson02}.

A visibility file of an individual pointing typically contains
$10^5 - 10^6$ visibility points. Thus, we need to bin the data into cells of
certain size prior to the Gaussianity analysis. 
For this paper, we have adopted the same bin size and binning procedure 
used for the determination of the power spectrum (see \cite{vsa3} and 
\cite{vsa5}). 
Thus, the compact array data were binned using
a cell size of $4\lambda$, where $\lambda$ is the observing wavelength, 
whereas for the extended array we used $9\lambda$. 
Tests with different cell sizes were done in A05, showing that there are no
significant changes in the results when varying these values.

\subsection{Gaussian simulations}

To illustrate the method in the case of mosaiced observations, we
used the VSA1 mosaic with the extended configuration \citep{vsa5}.
This mosaic contains three individual pointings (VSA1E, VSA1F and VSA1G).
After binning using $9\lambda$ cells, the data files contain 914, 882 and 911
complex visibility points, respectively. 
Thus, the covariance matrix has in this case a size of $N_d = 5414$ 
($N=2707$). 

We performed $10 000$ Gaussian simulations of this three-pointings mosaic, 
including Gaussian CMB signal plus Gaussian noise, according to the
noise levels of the real observations. 
The power spectrum adopted for the simulations, as well as for their
analysis, corresponds to the one presented in \cite{vsa7}.

\begin{figure*}
\begin{center}
\includegraphics[width=\columnwidth]{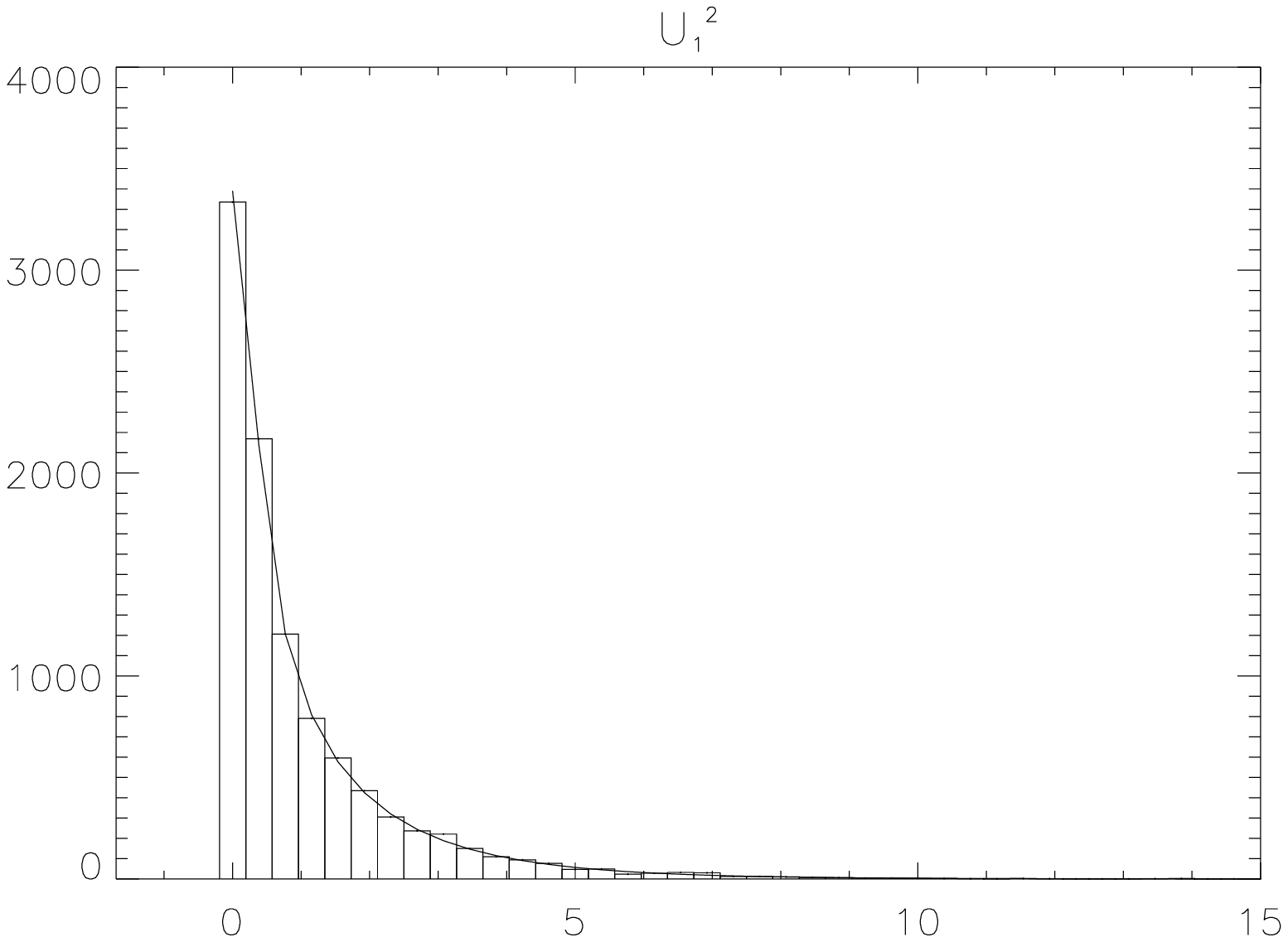}%
\includegraphics[width=\columnwidth]{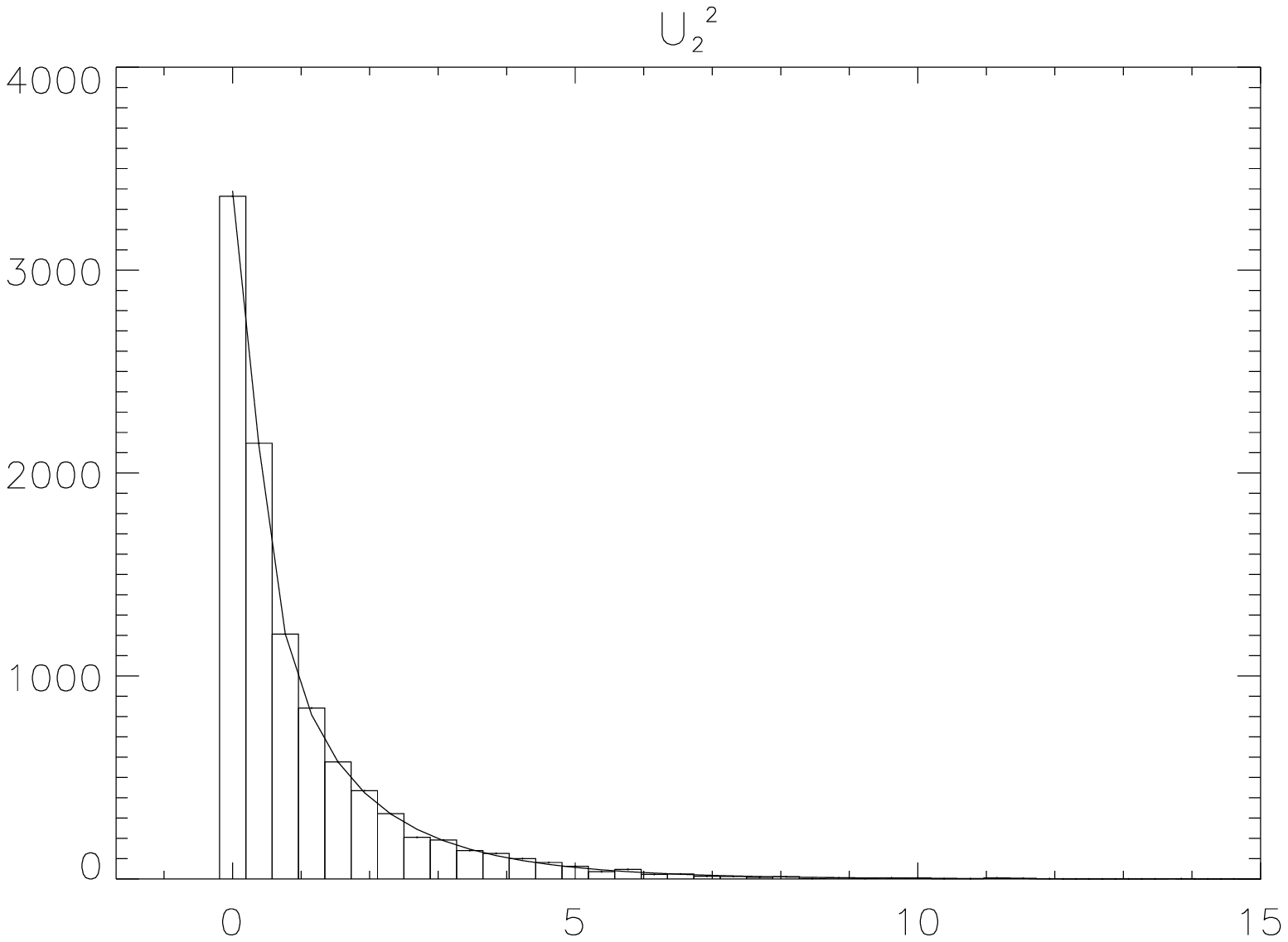}
\includegraphics[width=\columnwidth]{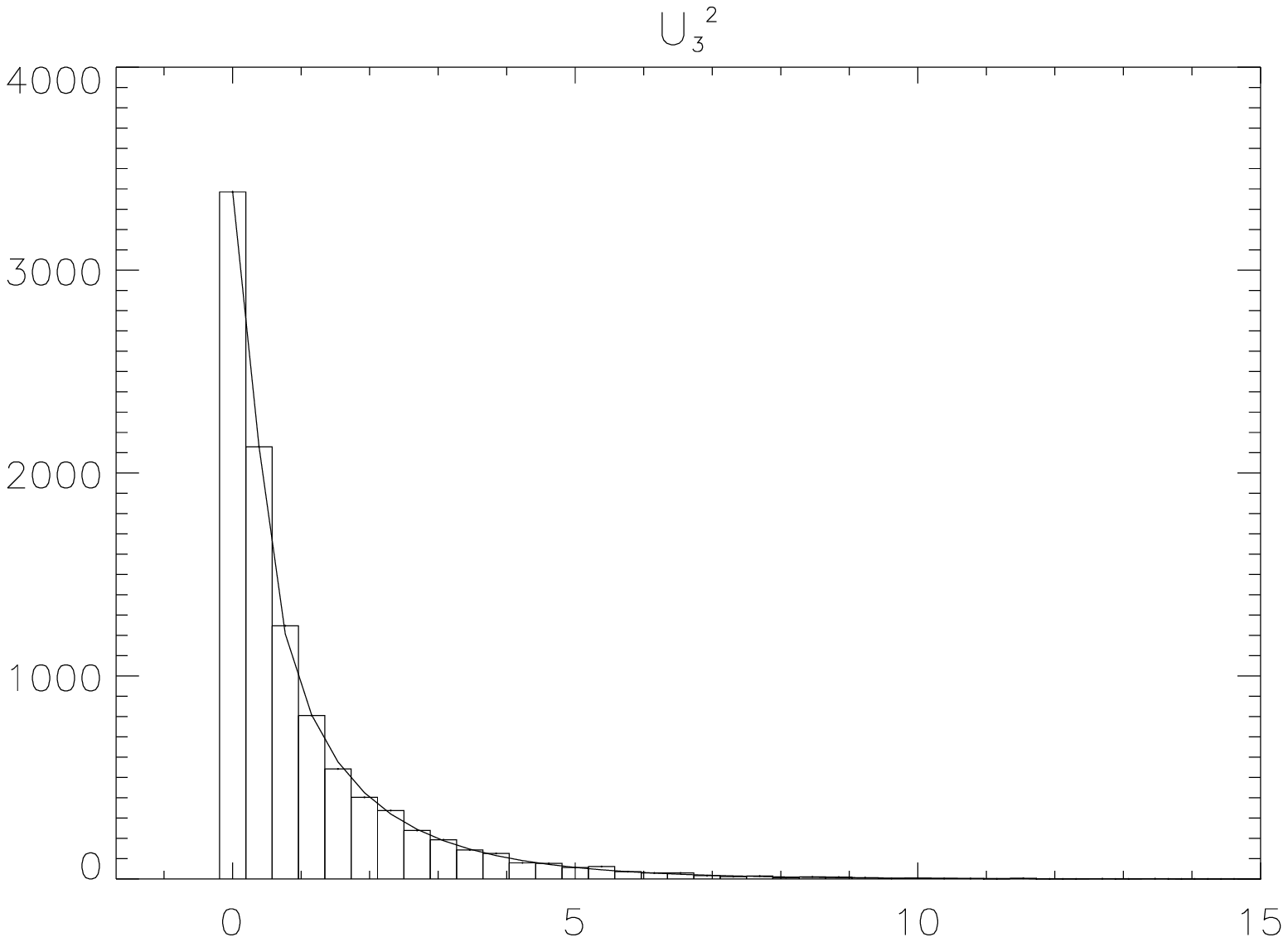}%
\includegraphics[width=\columnwidth]{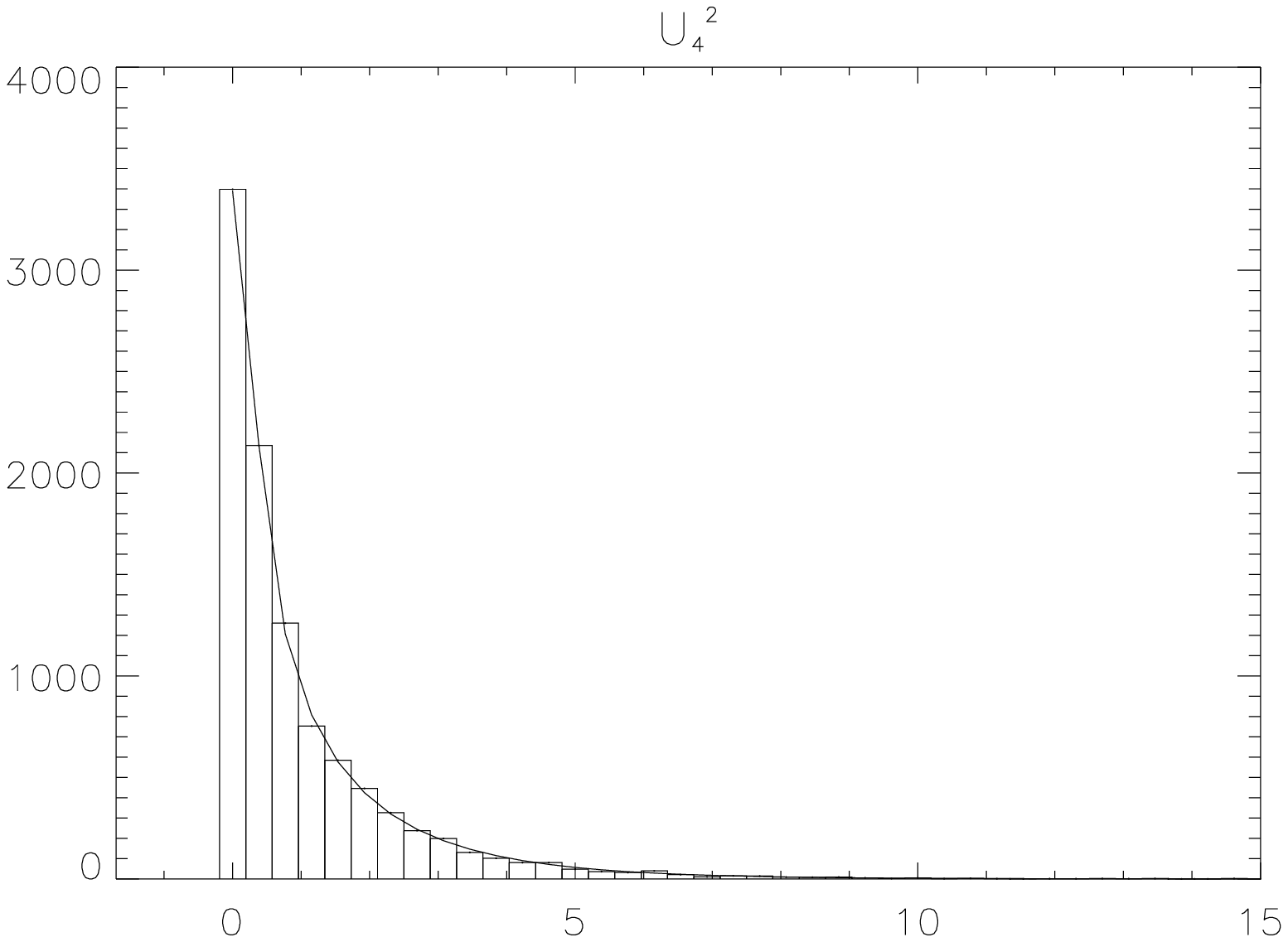}
\caption{Distributions of the $U_i^2$ statistics, from left to right, top to
  bottom $i=1,2,3,4$, when using only $E_i > 0.4$. 
They are obtained from $10 000$ simulated observations of
  the VSA1 mosaic, so each simulation contains three individual pointings which
 partially overlap on sky. The simulations contain Gaussian CMB signal plus
the realistic noise achieved in the observations. Prior to the analysis, the
  visibilities are binned in cells of $9\lambda$ in Fourier space. The solid
  line shows the expected values from a $\chi_1^2$ distribution normalised to
  the total number of simulations. }
\label{fig:chi2}
\end{center}
\end{figure*}

Fig.~\ref{fig:chi2} shows the histogram of the obtained $U_i^2$ values, and
compares it with the expected $\chi_1^2$ distribution. 
As expected, the $U_i^2$ quantities are distributed following a $\chi_1^2$.
Table~\ref{tab:u_i} presents the mean value and the standard deviation
of these $U_i^2$ quantities. 
As we will see below, we will focus our analysis on those values for the 
statistics which are computed using a subset of the signal-to-noise 
eigenmodes with high signal-to-noise ratios, i.e. 
using only those eigenmodes with associated eigenvalues satisfying 
$E_i \ge E_{cut}$. In particular, we will use the value of 
$E_{cut}=0.4$, so we present here the results both for $E_{cut}=0$
and $E_{cut}=0.4$. In this second case, we keep only $\sim 4\%$ of the data 
($219$ points), so this is why the distribution of $U_4^2$ is slightly
broader than the asymptotic value of $\sqrt{2}$.

\begin{table}
\begin{center}
\caption{Values of the mean ($\langle U_i^2 \rangle$) and the standard 
deviation ($\sigma$) of the statistics $U_i^2$ for 10000 Gaussian CMB plus 
noise simulations of the VSA1 extended mosaic. They are compared with the 
corresponding asymptotic values (for $\chi_1^2$), displayed in last column.
We show the results for two different values of $E_{cut}$. The numbers within
parenthesis indicate the number of eigenvalues with $E_i \ge E_{cut}$. }
\begin{tabular}{c c c c c c }
\hline
\hline
          & $U_1^2$ & $U_2^2$ & $U_3^2$ & $U_4^2$ & $\chi^2_1$  \cr
\hline
\multicolumn{6}{c}{$E_{cut}=0$ ($5414$ data)} \cr
\hline
$\langle U_i^2 \rangle$             
           & 1.01  & 0.99  &  1.00 & 0.99  &   1.00   \cr

$\sigma$   & 1.42  & 1.39  &  1.40 & 1.41  &   $\sqrt{2}$   \cr
\hline
\multicolumn{6}{c}{$E_{cut}=0.4$ ($219$ data) } \cr
\hline
$\langle U_i^2 \rangle$             
           & 1.00  & 1.02  &  0.98 & 0.99  &   1.00   \cr

$\sigma$   & 1.38  & 1.42  &  1.47 & 2.03  &   $\sqrt{2}$   \cr
\hline
\hline
\end{tabular}
\label{tab:u_i}
\end{center}
\end{table}

\begin{figure}
\begin{center}
\includegraphics[width=0.75\columnwidth,angle=90]{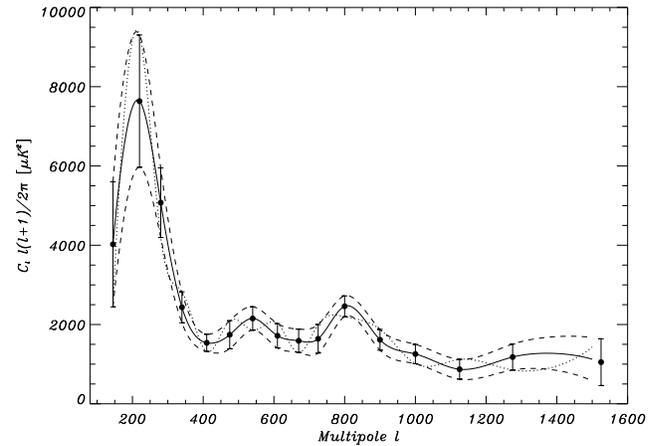}
\caption{Power spectrum used in the simulations. The data points correspond
to the data presented in Dickinson et al.(2004), 
but no correction due to residual sources
and Galactic foregrounds have been applied. 
The solid line is a spline interpolation of the data points, 
whereas the dashed lines are obtained interpolating 
the regions of $+1$ sigma and $-1$ sigma. 
The dotted line corresponds to an intermediate case, where
we connect alternate values of plus and minus one sigma. }
\label{fig:ps} 
\end{center}
\end{figure}

In order to illustrate the sensitivity of the method to the power spectrum
used in the computation of the covariance matrix, we have done the following
test. Using the measured power spectrum from \cite{vsa7}, we have created 
three mock power spectra, two of them defined by the envelope of 
the 1-sigma error bars of the data, and the third one
as an intermediate case connecting alternate values 
of $+1$ and $-1$ sigma, as shown in Figure~\ref{fig:ps}. 
 We call the ``upper power spectrum'' 
(``lower power spectrum'')  the one derived linking the  
measured data points plus (minus) one sigma, while the ``oscillating 
power spectrum'' is the one derived linking the alternate  
measured data points plus/minus one sigma.  

We use these power spectra to analyse the previous $10 000$ simulations
(which were generated using the measured power spectrum). The
mean values and the standard deviations
of the $U_i^2$ quantities are shown in Tables~\ref{tab:u_i_1}, 
\ref{tab:u_i_2} and \ref{tab:u_i_3}, respectively. 
These test cases show us how the distribution of the 
$U_i^2$ quantities is changing due to the use of an incorrect power spectrum.
Overall deviations of the power with respect to the true power spectrum
always appear as an excess in the $U_2^2$ statistic. 
Therefore, such an excess in $U_2^2$
can reflect either an intrinsic non-Gaussianity or a deviation
of the local power spectrum from the averaged one (i.e. anisotropy).
However, the third test case 
shows that if our power spectrum deviates from the real one
in a ``realistic way'', then we do not expect a significant effect 
on the mean value of $U_2^2$ 
statistic\footnote{However, we note that for this particular case 
in which the "wrong" power spectrum oscillates around the real one, 
we find an effect in the mean value of 
$U_4^2$ for $E_{cut}=0.4$, where we have $<U_4^2>=0.90$ with 
an error $1.69 / \sqrt{10000} \approx 0.02$. This shows that 
although the average band power is approximately the same as the true value (and
thus we do not detect a significant effect on $U_2^2$), the different shape of 
the power spectra inside each bin is detected in the higher moments ($U_4^2$).}. 
This is an important point, because a too strong dependence on the input 
power spectrum would make this method difficult to apply in practice.

\begin{table}
\begin{center}
\caption{Same as Table~\ref{tab:u_i}, but now the 10000 simulations are
analysed using the lower fit power spectrum explained in the text. }
\begin{tabular}{c c c c c c }
\hline
\hline
          & $U_1^2$ & $U_2^2$ & $U_3^2$ & $U_4^2$ & $\chi^2_1$  \cr
\hline
\multicolumn{6}{c}{$E_{cut}=0$ ($5414$ data)} \cr
\hline
$\langle U_i^2 \rangle$             
           & 0.98  & 1.18  &  1.00 & 1.04  &   1.00   \cr

$\sigma$   & 1.37  & 1.67  &  1.41 & 1.59  &   $\sqrt{2}$   \cr
\hline
\multicolumn{6}{c}{$E_{cut}=0.4$ ($198$ data)} \cr
\hline
$\langle U_i^2 \rangle$             
           & 1.13  & 2.44  &  1.37 & 1.56  &   1.00   \cr

$\sigma$   & 1.57  & 3.24  &  2.18 & 3.77  &   $\sqrt{2}$   \cr
\hline
\hline
\end{tabular}
\label{tab:u_i_1}
\end{center}
\end{table}

\begin{table}
\begin{center}
\caption{Same as Table~\ref{tab:u_i}, but now the 10000 simulations are
analysed using the upper fit power spectrum explained in the text.}
\begin{tabular}{c c c c c c }
\hline
\hline
          & $U_1^2$ & $U_2^2$ & $U_3^2$ & $U_4^2$ & $\chi^2_1$  \cr
\hline
\multicolumn{6}{c}{$E_{cut}=0$ ($5414$ data)} \cr
\hline
$\langle U_i^2 \rangle$             
           & 0.99  & 1.08  &  0.98 & 0.96  &   1.00   \cr

$\sigma$   & 1.37  & 1.67  &  1.41 & 1.59  &   $\sqrt{2}$   \cr
\hline
\multicolumn{6}{c}{$E_{cut}=0.4$ ($243$ data)} \cr
\hline
$\langle U_i^2 \rangle$             
           & 0.89  & 1.79  &  0.75 & 0.71  &   1.00   \cr

$\sigma$   & 1.44  & 1.49  &  1.39 & 1.47  &   $\sqrt{2}$   \cr
\hline
\hline
\end{tabular}
\label{tab:u_i_2}
\end{center}
\end{table}

\begin{table}
\begin{center}
\caption{Same as Table~\ref{tab:u_i}, but now the 10000 simulations are
analysed using the oscillating power spectrum explained in the text.}
\begin{tabular}{c c c c c c }
\hline
\hline
          & $U_1^2$ & $U_2^2$ & $U_3^2$ & $U_4^2$ & $\chi^2_1$  \cr
\hline
\multicolumn{6}{c}{$E_{cut}=0$ ($5414$ data)} \cr
\hline
$\langle U_i^2 \rangle$             
           & 0.99  & 0.99  &  1.02 & 1.00  &   1.00   \cr

$\sigma$   & 1.44  & 1.39  &  1.41 & 1.44  &   $\sqrt{2}$   \cr
\hline
\multicolumn{6}{c}{$E_{cut}=0.4$ ($220$ data)} \cr
\hline
$\langle U_i^2 \rangle$             
           & 0.97  & 1.01  &  0.96 & 0.90  &   1.00   \cr

$\sigma$   & 1.35  & 1.40  &  1.46 & 1.69  &   $\sqrt{2}$   \cr
\hline
\hline
\end{tabular}
\label{tab:u_i_3}
\end{center}
\end{table}

\subsection{Example of application of the Arnoldi method to VSA data}

We show now an example of application of the Arnoldi method to the 
VSA data, and we present the result of the analysis of the 
VSA2 mosaic observed in the first campaign of
the extended configuration. We have performed this analysis using both 
the standard method and the Arnoldi method. 

This mosaic is built up from three individual
pointings, with names VSA2E, VSA2F and VSA2G. 
The total number of visibility points for this mosaic, 
once the data are binned into $9\lambda$ cells, is $2751$, so the
dimension of the matrix $\bmath{A}$ in this case is $p=5502$.
We have considered here the case of $m=1000$.

\begin{figure}
\begin{center}
\includegraphics[width=\columnwidth]{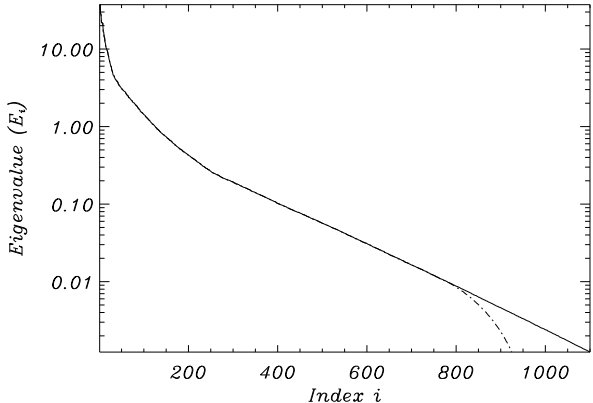}
\caption{Example of application of the Arnoldi method. We show the eigenvalues
  corresponding to the analysis of a mosaiced field with $5502$ data points. We
  have applied the Lanczos implementation using $m=1000$. 
  The solid line shows the eigenvalues ($E_i$) of $\bmath{A}$ derived from the
  analysis of the full covariance matrix, while the dot-dashed line shows the 
  eigenvalues ($E_i$) of $\bmath{H}$ from the Arnoldi method. }
\label{fig:lanczos} 
\end{center}
\end{figure}

Figure~\ref{fig:lanczos} shows the comparison of the eigenvalues derived 
using both methods. The eigenvalues of $\bmath{A}$ are represented by the solid 
line, while the eigenvalues of $\bmath{H}$ are displayed by the 
dot-dashed line.
We can see that the highest eigenvalues are recovered very well, but when
we approach to the dimension of 
the $\bmath{H}$ matrix (m=1000), then the recovered
eigenvalues departure from real values.
In this particular case, all eigenvalues with index $i<773$ are recovered
with a relative error smaller than $0.1\%$, while for $i<787$ the error is
smaller than $1\%$. In general, when the exact values of the eigenmodes are unknown, the
errors are controlled with the residual norm associated to the Ritz eigenvectors
\citep{saad}.

We note that for $E_{cut}=0.4$, 
the first $228$ eigenvalues are inside this cut. As we will see below,
we will use this value for the analysis. Therefore, for our purposes
an analysis using the Arnoldi method will provide exactly the same
results for the $U_i^2$ quantities as the full analysis.

\section{Gaussianity analysis of VSA data}

In this section, we present the results of the non-Gaussianity analysis of
the VSA data using the STGOF. 
The method has been applied to the list of pointings quoted in
Section~2, considering both the individual pointings separately, 
or arranging them into the corresponding mosaics.
For each analysis, the $U_i^2$ statistics were obtained 
for different values of $E_{cut}$, ranging from $0$ to $0.5$. 

In all cases (except for the Corona-Borealis mosaic), 
the analysis was performed using both
the standard procedure (full diagonalization of the covariance matrix) 
and the Lanczos algorithm with $m=1000$.   
We have checked that the values of the statistics ($U_i^2$) derived
from both methods are exactly the same for those cuts with $E_{cut} \ge 0.1$,
and there are small differences for $0.1 \ga E_{cut} \ga 0.01$. 
However, the standard computation was more time consuming than the 
Lanczos algorithm. For example, once the covariance matrix is computed, 
the typical computing time for the diagonalization of a matrix with $N_d=7200$ 
was $\approx 1.2$~hours in a 2.6~GHz Processor with 2~GHz RAM, and
this number scales roughly as $N_d^3$. 
However, the Lanczos algorithm with $m=1000$ takes only $15$~s in 
the diagonalization step.

The power spectrum used to compute the covariance matrices 
corresponds to the one derived from the data \citep{vsa7}, and presented
as a solid line in Fig.~\ref{fig:ps}. 
As mentioned above, the data files are binned in the visibility space
into cells of  $4\lambda$ for the compact array data, and $9\lambda$ 
for the extended array data.

Given the huge number of statistics we obtained, we proceed as follows
in order to present a comprehensive summary of the analysis. 
We shall present our results only 
for the analyses of the mosaiced observations, and
we shall quote the values and the significance of the 
$U_i^2$ statistics for the eigenvalue cut $E_{cut}=0.4$. 
High values of $E_{cut}$ are desirable because we 
select eigenmodes with higher signal-to-noise ratios, 
and for these values the statistical
properties of the signal are not diluted by the noise. However, if the value
of $E_{cut}$ is too high, then we end up with a small number of data-points.
As shown in A05, for the signal-to-noise ratio achieved in a VSA field, this 
cut for the eigenvalues uses a reasonable number of data points 
($\sim 10\%$), and at the same time it gives 
good results in discriminating non-Gaussian
signal obtained from the Edgeworth expansion and from string simulations.
In any case, we have checked that there are no significant differences if
we change the value of $E_{cut}$ in the range $0.1$ to $0.5$.

If a non-Gaussian signal is detected in a given mosaic, we follow these
steps

\begin{itemize}
\item[a)] We present the values of the statistics for the individual 
fields, in order to identify and localise 
the pointing(s) responsible of the non-Gaussian signal.
\item[b)] Data corresponding to those individual pointings containing the
non-Gaussian signal are split into two parts, corresponding to
different epochs of observation. The analysis is performed in each one of
these two parts, in order to isolate possible residual systematic effects.
\item[c)] Those individual pointings are also analysed by splitting the data
into separate regions of the uv-plane, so one can localise the origin of the
signal in Fourier space. To allow a simple identification of the Fourier
regions, we divided the uv-plane into $16$ concentric annuli. 
The edges of these 
annuli are taken from the bins adopted in \cite{vsa7} to present the power
spectrum results, and are quoted in Table~\ref{tab:bins} 
(note that $\ell = 2\pi |\bmath{u}|$).
\item[d)] The VSA collaboration has always maintained two independent pipelines, so every
pointing has been reduced in parallel by at least two of 
the three institutions. Thus, if a non-Gaussian signal is detected in 
a given mosaic, we also checked the second (independent) version of the 
reduction of the data, to identify if the non-Gaussian signal was due 
to residual systematic effects.
\item[e)] Finally, we have also explored the robustness of the 
results when using different noise estimates for the visibilities. 
Within the VSA collaboration, we have used two different noise estimates, one
based on daily estimates from the scatter on the visibility data in each
baseline, 
and another one based on the scatter of the visibilities 
when they are binned into cells prior to the power spectrum computations 
(see the details in \cite{vsa2}). 
These two methods have been shown to produce consistent results on the
power spectra, but we shall explore 
here whether those non-Gaussian signals could be understood 
using a different noise characterization. 
\end{itemize}

\begin{table}
\caption[]{Multipole bins considered in our analysis. The quoted values 
correspond to the same bin limits used in the power spectrum estimation,
as presented in \cite{vsa7}.}
\label{tab:bins}
\centering
\begin{tabular}{@{}ccc}
\hline
Bin & $\ell_{\rm min}$ & $\ell_{\rm max}$ \\
\hline
1  &        100   &      190 \\
2  &        190   &      250 \\
3  &        250   &      310 \\
4  &        310   &      370 \\
5  &        370   &      450 \\
6  &        450   &      500 \\
7  &        500   &      580 \\
8  &        580   &      640 \\
9  &        640   &      700 \\
10 &        700   &      750 \\
11 &        750   &      850 \\
12 &        850   &      950 \\
13 &        950   &     1050 \\
14 &       1050   &     1200 \\
15 &       1200   &     1350 \\
16 &       1350   &     1700 \\ 
\hline
\end{tabular}
\end{table}

Data has been split according to the configuration of the instrument
(compact or extended data). 
This will permit us to isolate possible systematic effects 
which could be only present in a given configuration.
For the extended array dataset, we also split the data into two subgroups, 
which correspond to the dataset presented in \cite{vsa5},
and the rest of the VSA extended fields described in \cite{vsa7}.
This allows a direct comparison of our results for the extended array
with those from the previous papers \citep{savage,smith} on the same datasets.

\begin{table*}
\caption{Values for the $U_i^2$ statistics and their corresponding probabilities
(within parenthesis) derived from the $\chi^2$ distribution 
from the analysis of the VSA mosaiced fields, 
using the cut $E_{cut}=0.4$. Second column shows the
number of visibility points ($N$) after binning the VSA data, prior to the 
Gaussianity analysis. The size of the covariance matrix is $N_d \times N_d$, 
with $N_d=2N$. The three cases where we have a non-Gaussian detection are 
marked using bold characters. }
\label{tab:results}
\centering
\begin{tabular}{@{}lccccc}
\hline
\hline
Mosaic & $N$ & $U_1^2$ & $U_2^2$ & $U_3^2$ & $U_4^2$  \\
\hline
VSA1 (compact)  & 4071 & 0.51 (52.6\%) & 0.61 (56.5\%) & 0.08 (21.9\%) & 1.45 (77.1\%) \\
VSA2 (compact)  & 3645 & 0.63 (57.4\%) & 2.43 (88.1\%) & 0.16 (30.7\%) & 0.00 (3.03\%) \\
VSA3 (compact)  & 4715 & 1.37 (75.9\%) & 0.19 (33.5\%) & 0.86 (64.7\%) & 1.27 (74.1\%) \\
\hline
VSA1 (extended I) & 2707 & 0.01 (7.06\%) & 2.40 (87.9\%) & 0.26 (39.2\%) & 0.07 (20.9\%) \\
VSA2 (extended I) & 2751 & 1.93 (83.5\%) & {\bf 8.50 (99.7\%)} & 0.22 (36.3\%) & 2.88 (91.0\%) \\
VSA3 (extended I) & 2925 & 0.56 (54.5\%) & {\bf 7.49 (99.4\%)} & 0.26 (38.8\%) & 0.39 (46.6\%) \\
\hline
VSA1 (extended II) & 4174 & 0.01 (7.62\%) & 0.37 (45.7\%) & 0.25 (38.0\%) & 0.03 (13.8\%) \\
VSA2 (extended II) & 4384 & 0.06 (19.0\%) & 1.34 (75.2\%) & 0.19 (33.5\%) & 1.53 (78.3\%) \\
VSA3 (extended II) & 4297 & 0.00 (0.46\%) & 2.23 (86.4\%) & 0.13 (27.8\%) & 0.00 (4.53\%) \\
VSA5 (extended II) & 2485 & 0.03 (13.3\%) & 2.26 (86.8\%) & 0.07 (21.1\%) & 0.16 (31.2\%) \\
VSA6 (extended II) & 2531 & 2.98 (91.6\%) & 1.13 (71.2\%) & 0.51 (52.5\%) & 0.37 (45.7\%) \\
VSA7 (extended II) & 2611 & 0.08 (22.9\%) & 0.16 (31.2\%) & 0.06 (20.0\%) & 0.00 (4.50\%) \\
VSA8 (extended II) & 2919 & 0.12 (27.4\%) & 0.58 (55.3\%) & 0.00 (2.17\%) & 0.06
(19.1\%) \\
\hline
CoronaB (extended II) & 6629 & 0.01 (6.21\%) & {\bf 9.79 (99.82\%)}  &  
0.26 (38.70\%) & 0.08 (21.75\%) \\
\hline
\hline
\end{tabular}
\end{table*}

Our results are presented in Table~\ref{tab:results}, for the case of
$E_{cut}=0.4$. 
The total number of visibility points in the binned files is also 
shown in the second column of that table.
All the results are compatible with a 
Gaussian distribution, except in three cases where we obtain values for
the statistics with a significance greater than $95\%$. 
These cases are the mosaics for 
the VSA2 and VSA3 fields obtained with the first release 
of the extended configuration (fields VSA2E, VSA2F and VSA2G, and
VSA3E, VSA3F and VSA3G, respectively), and the Corona-Borealis mosaic
(quoted as CoronaB). 
In these three cases, there is apparently a detection with the
$U_2^2$ statistic, which may indicate either a deviation from the theoretical
power spectrum or a non-Gaussian signature (as shown for instance for 
cosmic strings in A05). 
We now discuss each of these three cases in detail.

\subsection{VSA2 mosaic, extended configuration}

An individual analysis of the three fields of the mosaic 
shows that VSA2E is compatible with gaussianity ($U_2^2 = 2.46 (88.3\%)$), 
while the VSA2F ($U_2^2 = 10.3 (99.9\%)$) and 
VSA2G ($U_2^2 = 17.0 (100\%)$) present a strong deviation. 

If we now split the VSA2F and VSA2G datasets into two parts, 
corresponding to two separate epochs,  
we find that the non-Gaussian signal is present in both of them, and
for both pointings. This result suggests that the origin of the signal
is intrinsic to the data, because it can not be isolated in a
separated epoch.
In order to localise the origin of this signal in Fourier space, we 
performed the analysis using the $16$ anulii regions mentioned above, and 
defined in Table~\ref{tab:bins}. 
In order to keep a reasonable number of data points in this ``bin analysis'', we now
use a lower value for $E_{cut} (=0.1)$. 
The detailed analysis shows that the non-Gaussian 
signal in $U_2^2$ associated with 
VSA2F is localised in bins 9 and 11, 
while the non-Gaussianity associated to VSA2G comes from bins 9, 10 and 11.

The following step, in order to cross-check these results, is to use
the independent reduction of these two pointings that we produced  
in the collaboration. 
An analysis of this second version of the data shows that both VSA2F and
VSA2G present a deviation on $U_2^2$, although in the case of the VSA2F pointing
this deviation is smaller ($U_2^2 = 5.1 (97.6\%)$). 
We have also checked that these non-Gaussianity 
detections are robust against the two different schemes for noise estimation. 
Moreover, varying the noise estimates for the data 
within a $\pm 5\% $ does not change the results on the $U_2^2$ statistic. 

\subsubsection{Study of the local power spectrum}

The previous tests suggest that the non-Gaussian signal detected in the
data via the $U_2^2$ statistic could be real, and not due to systematics. 
Given that $U_2^2$ may indicate deviations on the second moment 
of the (transformed) visibilities, 
we have investigated the power spectrum of these two fields, as well as 
that of the VSA2EFG mosaic.

We have used for this computation $8$ bins instead of $16$ because
the thermal noise in a single VSA field is high, so using
small amount of visibilities could bias the estimation of the power. 
The $8$ bins were obtained by joining the sixteen
bins from Table~\ref{tab:bins} in pairs, so one can easily relate the new bins
with the old ones (i.e. the new bin 1 corresponds to bins 1 and 2 from that 
table, and so on).

\begin{figure}
\begin{center}
\includegraphics[width=0.75\columnwidth,angle=90]{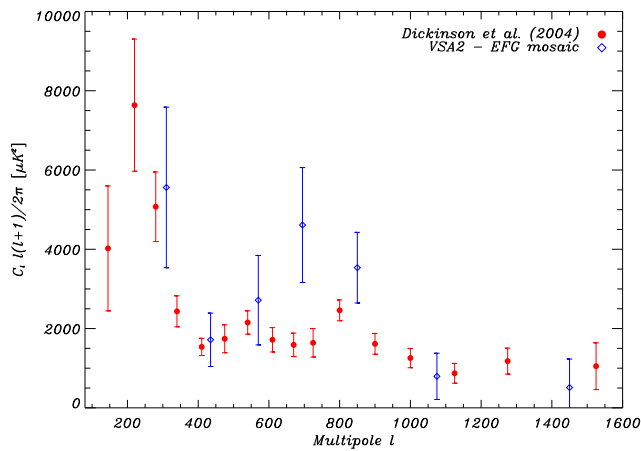}
\caption{Power spectrum of the VSA2 mosaic observed with the 
extended I configuration (pointings VSA2E, VSA2F and VSA2G). Although
noisy (there are only three pointings entering in the computation), we
can see a deviation with respect to the Dickinson et al. result, at
angular scales corresponding to $\ell \approx 700-900$. }
\label{fig:ps_vsa2efg} 
\end{center}
\end{figure}

The power spectrum of the VSA2EFG field is found to be 
compatible (at the 1-sigma level) with the one presented in 
\cite{vsa7}, except in two bins which correspond
with bins 9 \& 10, and 11 \& 12 from Table~\ref{tab:bins}, and where 
we find a $2.1$-sigma and $1.5$-sigma deviation toward higher values, 
respectively (see Fig.~\ref{fig:ps_vsa2efg}). Although noisy, 
the power spectrum of the individual fields (VSA2F and VSA2G) also show 
a deviation on those scales, being larger in the VSA2G case (practically $2$-sigma).
The VSA2E field shows no deviations. 
A visual inspection of the VSA2 extended I mosaic  (see figure~3 in 
Dickinson et al. 2004) shows an intense positive feature close to the 
centre of the G pointing, 
which could be the responsible for that deviation. 

In order to check if these fields are intrinsically Gaussian, we 
have analysed the VSA2EFG mosaic using now its own power spectrum.
In this case, we find the value $U_2^2 =  1.01 (68.38\%)$  for $E_{cut}=0.4$,
which is now compatible with Gaussianity. These results suggest that
the $U_2^2$ excess found in this case is connected with a deviation of the
power spectrum of the region with respect to the average one, and not with an
intrinsic non-Gaussianity (in the sense that 
when we perform the analysis using their power spectrum, then they are Gaussian).

To complete our study with the STGOF, we have investigated how
significant is that deviation of the power spectrum
of the region with respect to the average power spectrum. 
The probability of finding such a deviation in a multivariate Gaussian field can be easily 
derived by noting that the $8$ bins used in the power spectrum computation are
practically independent. Using Monte Carlo simulations, we estimate that 
with the measured band powers and errors, the 
probability of having $2$ numbers out of $8$ such that one of them deviates $2.1$-sigma and
the other one $1.5$-sigma is $4.6\%$. If we now impose that these two
values should be in adjacent bins (as is our case), then this value reduces to $1.4\%$.
Given that we have observed $13$ mosaics with similar characteristics (although not
completely independent, because mosaics 1, 2 and 3 for different configurations
partially overlap), then 
we conclude that this deviation of the power spectrum is not
as significant as the one obtained using the $U_2^2$ statistic.
In any case, both analyses suggest that we have detected a local
deviation of the power spectrum (i.e. an anisotropy) in this VSA2
mosaic.

\subsection{VSA3 mosaic, extended configuration}

An individual analysis of the three fields contained in the mosaic 
shows that VSA3F is compatible with gaussianity ($U_2^2 = 0.52 (52.8\%)$), 
while the VSA3E ($U_2^2 = 5.48 (98.1\%)$) and 
VSA3G ($U_2^2 = 5.89 (98.5\%)$) present a deviation. 

We proceed as in the previous case, and we first split the two datasets (VSA3E and VSA3G) into
two separate parts corresponding to different observation epochs. 
We find that the non-Gaussian signal connected with VSA3G is only present in 
one part of the data, while the one corresponding 
to VSA3E is absent in both parts. 
This suggests that at least the non-Gaussian signal 
found in VSA3G with $U_2^2$ could be due to systematic effects, 
because is only present in one part of the data. 

Next, we have examined the results using the two different 
noise estimation schemes. Again, varying the noise estimates within 
$\pm 5\%$ does not change the results on $U_2^2$. 
However, in this case we find a difference between the two methods for the
VSA3E and VSA3G pointings. When using a noise estimation based on 
the scatter of the visibilities when binning all daily observations, 
we find that 
now VSA3G is compatible with Gaussianity ($U_2^2 = 0.33 (43.4\%)$), and
VSA3E shows a marginal deviation ($U_2^2 = 3.85 (95.0\%)$).
These results suggest that
the non-Gaussian signal found in this mosaic is in reality produced by
residual systematic effects associated with a few visibility points which
were left in the analysis.
This is confirmed when using 
the other independent data reduction of these two pointings, 
where we find that both VSA3E and VSA3G are compatible with Gaussianity. 

The detailed analysis in separate bins permits us to isolate the origin
of this signal, and exclude the affected visibilities. 
In both cases, it is connected with one single bin (5 and 11, respectively).
Once these few visibilities are removed, all 
the individual pointings become compatible
with Gaussianity in the two versions of the data reduction, and 
the joint analysis of the three-fields mosaic gives $U_2^2 = 1.85 (82.7\%)$.

\subsection{Corona Borealis mosaic, extended configuration}

The previous analysis of the VSA data shows that the STGOF tests
are very sensitive methods to detect residual systematic effects in the data
and/or deviations of the power spectrum from the average one. 
To further check the power of this method, 
we have also applied it to the analysis of the data from a survey in the 
Corona Borealis supercluster region with the VSA \citep{ricardo}. 
These data are known to present a strong deviation from Gaussianity,
associated to a negative decrement in the map which can not be
explained in terms of primordial CMB fluctuations, 
or associated to a (known) cluster of galaxies in the region. 
A power spectrum analysis of these data presents a clear deviation 
with respect to the cosmological one in angular scales around 
$\ell \approx 500$, which corresponds to the angular size of 
that negative decrement.

Here, we will complete the Gaussianity studies described in \cite{ricardo} by 
performing the simultaneous analysis of the data with the STGOF.
We have analysed the $9$ pointings which altogether make a mosaic of
the central region of the supercluster. Those pointings are noted
with letters A,B,C,D,E,H,I,J and K. 
Given the large size of the full covariance matrix for this dataset ($N_d = 13258$), only
the Lanczos method was applied. As shown above, for the
signal-to-noise ratios achieved in the VSA observations, the Lanczos
method with $m=1000$, when applied to covariance matrices with $N_d \sim
7000$, gives the same values of the $U_i^2$ as those
obtained with the full analysis if we restrict ourselves to $E_{cut}
\ga 0.01$ (i.e. $77~\%$ of the $1000$ eigenvalues are good
approximations to the true values). In the Corona Borealis case, only
$\sim 330$ eigenvalues are found to be above the cut $E_{cut}=0.4$, so 
the use of the Lanczos algorithm is justified. 

In the last row of Table~\ref{tab:results} 
we present the results obtained with the Arnoldi algorithm 
considering $m=1000$ and $E_{cut}=0.4$. 
These numbers show a deviation of the $U_2^2$ statistic, as we would
expect from the fact that the power spectrum of this mosaic differs from the
average one.

We now proceed as in the previous cases, and 
we first analyse all the individual
pointings which participate in the mosaic. 
Considered as a whole, each one of the
$9$ pointings seem to be consistent with Gaussianity, 
although the highest $U_2^2$ 
value is found for the H pointing ($U_2^2= 3.72 (94.6\%)$) 
which contains the main decrement near its centre. 
However, a detailed analysis in separate bins of all 
the nine datasets shows that
the H pointing is the only one presenting a strong 
deviation for $U_2^2$, and which
is associated with the multipole region around 
$\ell \approx 500$, as we would expect.  
These results are stable when splitting the data 
in two separate parts, so the non-Gaussianity
seems to be intrinsic to the data. The results are also the
same for the two noise estimation schemes. 

Finally, we have re-analysed the whole 
Corona-Borealis mosaic using its own 
power spectrum, in order to probe if the detected 
non-Gaussianity is only associated
to the fact that we have a deviation of the power spectrum.
In this case, we obtain that $U_2^2 = 7.44 (99.4\%)$. This result is
very interesting, because it is showing that in this case, 
the non-Gaussianity found in the Corona-Borealis mosaic is intrinsic, and 
is not associated to a deviation of the power spectrum: 
even when we use the correct (local) power spectrum of the region in the analysis, the
$U_2^2$ is still showing a detection of a non-Gaussianity.

\section{Discussion and conclusions}

We have analysed the full VSA data sets presented in \cite{vsa2}, 
\cite{vsa5}, \cite{vsa7} and \cite{ricardo},
using the Smooth Tests of Goodness-of-Fit adapted to interferometer experiments.
This method was described in A05, but here it has been extended to deal
with large mosaics via the Arnoldi method. This numerical method permits to
solve large eigenvalue problems by reducing the dimensionality 
of the covariance matrix. We have shown that one implementation of this
method for Hermitian matrices, the Lanczos algorithm, is able to provide good 
approximations to those eigenvalues and eigenmodes of the 
full covariance matrix with larger signal-to-noise ratio.

From our analysis of the VSA data dedicated to cosmological studies, 
we found that out of the $13$ mosaics presented in Table~\ref{tab:results}, 
eleven of them are consistent with Gaussianity, and two of them 
show a deviation from Gaussianity. 
In one case (mosaic VSA3E + VSA3F + VSA3G) the non-Gaussian signal is
shown to be produced by few visibility points which contain systematic
effects that were not removed properly from the data. Once this data 
are removed, the mosaic becomes compatible with Gaussianity.

In the second case (mosaic VSA2E + VSA2F + VSA2G), we show that the 
method is detecting a local deviation of the
power spectrum with respect to the average one. 
The STGOF are very sensitive to the power spectrum
adopted for the computation of the signal-to-noise eigenmodes, so
small deviations from the correct power spectrum are easily detected 
in the analysis. This ability of the method could be used to study the 
isotropy of a given CMB map. 
Moreover, when the data of this mosaic are analysed
with their (local) power spectrum, then they are compatible with
Gaussianity; but when we analyse the whole data set with 
the local power spectrum of the VSA2EFG mosaic, then the rest of 
the mosaics become incompatible with Gaussianity. 
These results could indicate the presence of anisotropy. 

However, there could be other possible explanations, such as the 
presence of residual foregrounds in this particular mosaic. 
Nevertheless, there are no significant features seen in multi-frequency 
foregrounds maps ($408$~MHz, H$\alpha$, 100~$\mu m$ dust map) 
that align with the main features of the VSA mosaic. 
Moreover, this mosaic is one of the cleanest regions (in terms of rms), as
shown in \cite{vsa2}. 
Regarding the case of point sources, 
we have investigated the possibility that this
deviation could be produced by two unsubtracted sources with fluxes around 
$40$~mJy. Note that a population of unresolved sources is not a 
possible explanation, because its contribution should 
scale as $\ell^2$ and it would produce  
deviations on smaller scales as well.
However, two sources would produce structure as a function of $\ell$. 
The value of $40$~mJy is an extrapolation to 33~GHz of the
completeness limit of the Ryle telescope survey at 15~GHz (see 
\cite{waldram}, and also \cite{kieran} for the details of the survey).
The $5$-sigma limit from Ryle Telescope at $15$~GHz is
10 mJy; the worst case that we can have is a rising spectrum 
source with index of $2$, so we considered the case of two $40$~mJy sources. 
Using Monte-Carlo simulations to explore all the possible relative
spatial distributions of the two sources, we find that we can not explain
the measured value for $U_2^2$ in the VSA2G.

It is also interesting to mention that the VSA2F and VSA2G fields
are practically contained within the VSA2 mosaic obtained with the compact
array. Although those data were obtained with a different
configuration, the scales where we find a deviation with the extended
array were also sampled with the compact array (but with poorer
signal-to-noise ratio), so we could expect a small signature in the
analysis. However, the value for VSA2 compact is
$U_2^2=2.43 (88.1\%)$, which is somewhat high but still compatible with
Gaussianity. 

We also note that the VSA2 and VSA3 fields were already analysed with
other Gaussianity tools in \cite{savage} and \cite{smith}, but no 
evidence of non-Gaussian
signals was found. This shows the importance of applying a wide
number of Gaussianity tests to the data, given that each particular
test is sensitive to a different type of non-Gaussianity. 

The fraction of data affected by the systematic effects in VSA3
extended I mosaic is too small to affect the published power spectrum
by the VSA collaboration. A re-evaluation of the complete VSA power spectrum
when we use the corrected version for these data
shows no differences with the published values (within the numerical
precision of the maximum likelihood code).
However, the deviation in the VSA2 extended I 
mosaic could influence these numbers. Although there is no 
justified reason to exclude this VSA2 mosaic from the final computation, 
we have quantified the effect of excluding those data from the final
analysis. 
Thus, we have considered the extreme case
in which we remove from the dataset all visibilities 
which lie inside one of the bins showing the non-Gaussianity. 
With this new dataset, we have re-evaluated the complete power 
spectrum, and compared it to the one presented in \cite{vsa7} in order
to obtain the maximum deviations that we would expect. 
Differences (within the numerical precision of the power spectrum
code) are only found in three bins (10, 11 \& 12), and are of the
order of $-9.2\%$, $-4.0\%$ and $+4.7\%$ with respect to the published
values. To complete this check, 
we have repeated the parameter estimation analysis described in
\cite{vsa8}, but now using this new power spectrum. 
We have considered two different models, corresponding to
use VSA+WMAP data on one hand, and VSA+COBE data on the other hand, and
we have explored the 6-parameter flat $\Lambda$CDM model described 
in Table~2 of \cite{vsa8}. 
The differences in all parameters for the VSA+WMAP case 
are found to be at the most 2\%, so we can conclude that the
cosmological analyses based on the published data are not affected. 
However, when using only the VSA+COBE data, we find only a significant 
difference 
in the estimate of the baryon density, which turns out to be 
$\Omega_b h^2 = 0.030^{+0.007}_{-0.005}$ at 68\% C.L., and which should be
compared with the former value $\Omega_b h^2 = 0.033^{+0.007}_{-0.007}$. 
This change can be easily 
understood, because the new power spectrum has less power in the
region of the third acoustic peak, giving a smaller (but compatible) value of 
the baryon density. Note that the new value is now closer to the BBN result,
as well as to the result of the analysis using WMAP+VSA. 

Finally, we have applied the method to VSA observations of the Corona Borealis
supercluster. These observations are known to present a deviation
on the power spectrum produced by a strong decrement in one of the 
pointings (see \cite{ricardo}). Our method is able to
detect this deviation, and it finds a large value for the $U_2^2$
statistic which can not be interpreted as systematic effects. 
A careful analysis shows that the non-Gaussian signal is associated
with the same scales where we find a deviation on the local power
spectrum, and which correspond to the angular scale of the negative decrement.
However, in this case the non-Gaussian signal detected by the method
is intrinsic to the data, in the sense that if we use the local
power spectrum and we repeat the analysis, a non-Gaussian detection is 
still present.

\section*{Acknowledgments} 
We thank the staff of the Mullard Radio Astronomy Observatory, the Jodrell Bank
Observatory and the Teide Observatory for invaluable assistance in the
commissioning and operation of the VSA. The VSA is supported by PPARC and the
IAC. Partial finantial support was provided by the Spanish Ministry of Science
and Technology project AYA2001-1657. We also acknowledge the Spanish project
ESP2004-07067-C03-01.

\label{lastpage}
\end{document}